\documentclass[traditabstract]{aa}
\usepackage{txfonts}
\usepackage{natbib}
\usepackage{graphicx}

\newcommand{\lta}{\lesssim}
\newcommand{\gta}{\gtrsim}

\newcommand{\kms}{\>{\rm km}\,{\rm s}^{-1}}
\newcommand{\kkms}{\>{\rm K}\,{\rm km}\,{\rm s}^{-1}}
\newcommand{\kkk}{{\rm K}\,{\rm km}\,{\rm s}^{-1}}

\newcommand{\kk}{\>{\rm K}}

\newcommand{\pc}{\>{\rm pc}}

\newcommand{\m}{\>{\rm m}}

\newcommand{\cm}{\>{\rm cm}}
\newcommand{\mm}{\>{\rm mm}}
\newcommand{\mum}{\>{\mu {\rm m}}}

\newcommand{\mhz}{\>{\rm MHz}}
\newcommand{\ghz}{\>{\rm GHz}}

\newcommand{\msun}{\>{\rm M_{\odot}}}

\newcommand{\as}{^{\prime\prime}}
\newcommand{\am}{^{\prime}}
\newcommand{\bdm}{\begin{displaymath}}
\newcommand{\edm}{\end{displaymath}}
\newcommand{\beq}{\begin{equation}}
\newcommand{\eeq}{\end{equation}}
\newcommand{\bit}{\begin{itemize}}
\newcommand{\eit}{\end{itemize}}
\newcommand{\ben}{\begin{enumerate}}
\newcommand{\een}{\end{enumerate}}
\newcommand{\bfi}{\begin{figure}[htb]}
\newcommand{\bpfi}{\begin{figure}[p]}

\newcommand{\htwo}{$\rm H_2$}
\newcommand{\coone}{$\rm ^{12}CO(1-0)$}
\newcommand{\cotwo}{$\rm ^{12}CO(2-1)$}
\newcommand{\ione}{$I_{10}$}
\newcommand{\itwo}{$I_{21}$}
\newcommand{\ihcn}{$I_{HCN}$}
\newcommand{\ihcoplus}{$I_{HCO+}$}
\newcommand{\itco}{$I_{{^{13}}CO}$}

\newcommand{\mhtwo}{$M_{\rm H_2}$}

\begin{document}

\title{Molecular gas around low-luminosity AGN in late-type spirals}

\author{Torsten B\"oker\inst{1}
  \and Eva Schinnerer\inst{2}
   \and Ute Lisenfeld\inst{3}}


\institute{Research and Scientific Support Department, European Space Agency, Keplerlaan 1, 2200 AG Noordwijk, The Netherlands
  \and Max-Planck-Institute for Astronomy, K\"onigstuhl 17, 69117 Heidelberg, Germany
  \and Departamento de F\'\i sica Te\'orica y del Comsom, Facultad de Ciencias, Universidad de Granada, Spain}

\date{Received  16 May 2011 / Accepted 26 July 2011}

\abstract{We have studied the molecular gas in the vicinity of low-luminosity active galactic
nuclei (AGNs) in 
three bulge-less spiral galaxies: NGC\,1042, NGC\,4178, and NGC\,4395. The (1-0) and (2-1) transitions of
gaseous carbon monoxide (CO) are clearly detected within the central kpc of all three galaxies.
In the case of NGC\,4395, this constitutes the first reported detection of CO. In general, the CO 
emission is faint, as may be expected from their less-than-spectacular star formation activity. 
Interestingly, however, both face-on galaxies in our sample 
(which allow an unimpeded view of their nucleus) show an elevated intensity 
ratio CO(2-1)/CO(1-0) when compared to similar late-type spirals {\it without} 
an AGN. We discuss that this is unlikely due to a very compact CO source. Instead, we 
speculate that even energetically weak AGN can impact the physical state of the surrounding gas. 
We do not detect any tracers of dense molecular gas such as HCN or HCO+, but the sensitivity
of our observations allows us to establish upper limits that lie at the low end of the range
observed in more energetic AGN. The derived gas density is less than
$n_{\rm H_2}\approx 2\cdot 10^3\,\rm cm^{-3}$ which is significantly lower than in most
other nearby galaxies. The scarcity of dense gas suggests that the conditions for star formation 
are poor in these nuclei.
}
\maketitle

\section{Introduction}
\label{sec:intro}
Recent studies have demonstrated that supermassive black holes
(SMBHs) reside in the nuclei of most bulge-dominated galaxies, and
that the mass of the SMBH, M$_{\rm BH}$, is strongly correlated with various
properties of the host spheroid \citep[e.g.][]{gebhardt00,ferrarese00}. 
This discovery has launched numerous
speculations that the formation and evolution of galaxies and their
SMBHs are fundamentally linked, and that perhaps the presence of a
bulge is a necessary ingredient for a black hole to form and grow.
Indeed, M33, the most nearby example of a truly bulgeless disk galaxy
shows no evidence of a SMBH, and the upper limit on the black hole
mass (determined from stellar dynamics) is significantly below that
predicted by the M$_{\rm BH}$-$\sigma$ relation for early-type
galaxies \citep{gebhardt01}. On the other hand, the galaxy NGC\,4395 
shows no evidence for a bulge either, yet it has long been known
to harbor an active nucleus \citep{filippenko89}. Until
recently, however, this galaxy was the only known example of an
accreting black hole in a bulgeless disk, leaving open the possibility
that it is just an anomaly.

\begin{figure*}[ht]
\centering
\resizebox{0.9\hsize}{!}{\rotatebox{0}{\includegraphics{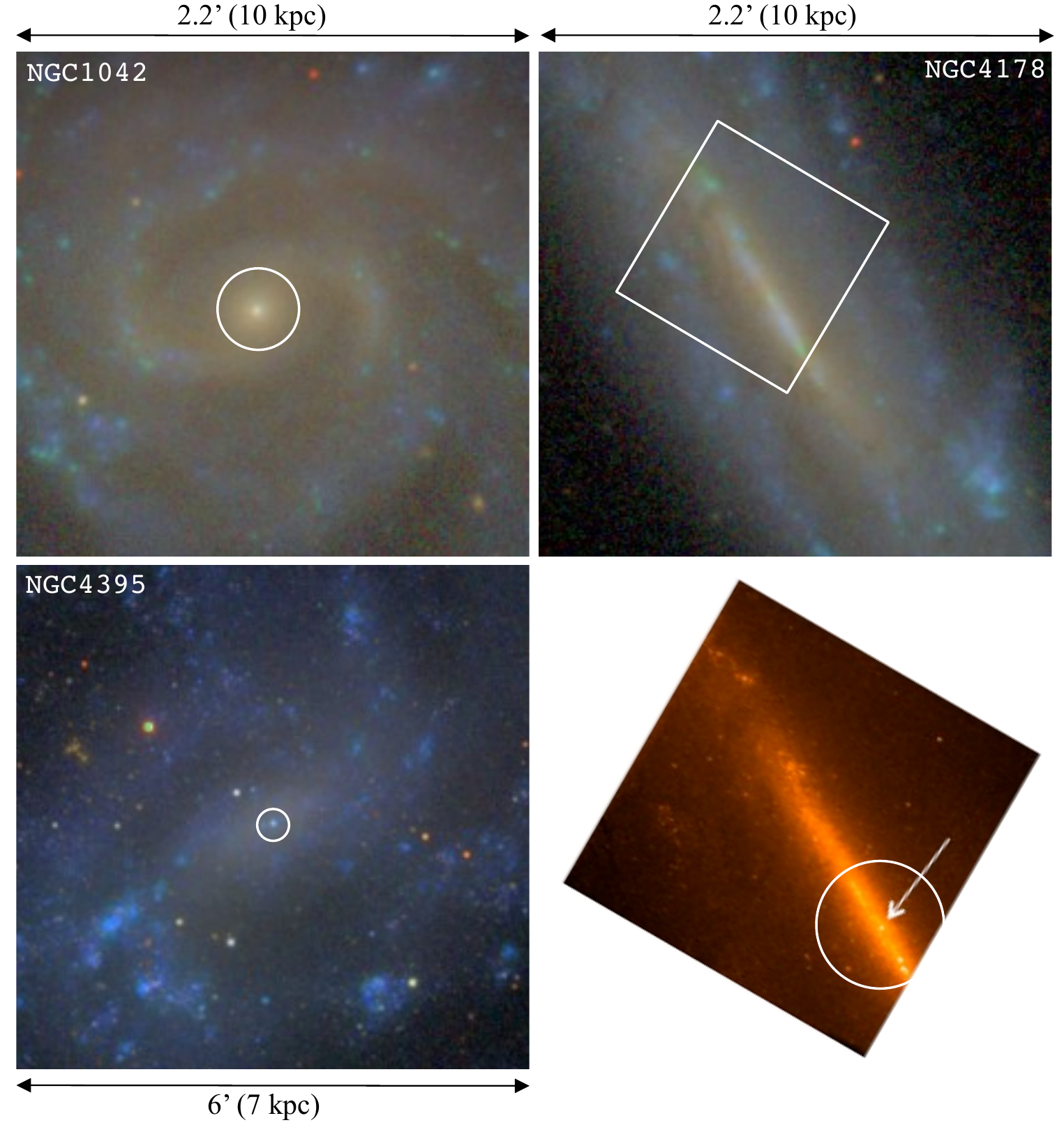}}} 
\caption{Optical/near-infrared images of our target galaxies. Approximate image
sizes are indicated, their orientation, however, is arbitrary. All three galaxies 
clearly harbor a nuclear star cluster (indicated by an arrow in the case
of NGC\,4178). The size and location of the IRAM beam used for the
observations described here ($21\as$ at 115\,GHz) is indicated by the circles. 
}
\label{fig:mosaic}
\end{figure*}

Over the past few years, however, a number of similar cases have been found
that have put into question whether black holes truly require a bulge
to form or grow.  These discoveries came from optical spectroscopy of
nuclear star clusters that revealed emission line ratios and profiles 
indicative of nuclear activity \citep[NGC\,1042;][]{shields08}, the detection
of high-excitation mid-infrared lines such as [NeV] 
\citep[NGC\,4178;][]{satyapal09}, or a combination of both 
\citep[NGC\,3621;][]{satyapal07,barth09}. In all cases, the AGN is
energetically weak, and the inferred SMBH masses are at the lower end
of the mass range probed by the established M$_{\rm BH}$-$\sigma$
relation, typically below $10^6\msun$. For these reasons, they are
often referred to as ``mini-AGN''.

While the number of such ``mini-AGN'' is likely to increase over the
coming years, it seems that they are indeed rare, i.e. the AGN
fraction is lower in ``pure'' disks than in more bulge-dominated
galaxies \citep{satyapal09}. A clue for why this may be so comes from the
recent finding that the nuclei of bulgeless disks are often occupied
by a compact, massive star cluster \citep[e.g.][]{boker02}. 
It has been speculated that the formation of such a nuclear star 
cluster (NSC) can inhibit the growth of a SMBH because gas is 
expelled from the nucleus by the mechanical feedback from massive
stars \citep{schinnerer08,nayakshin09}. While the relative importance of the SMBH
indeed appears to drop strongly towards lower spheroid masses,
it is interesting to note that all ``mini-AGN'' known so
far are located in galaxies {\it with} an NSC \citep{satyapal09}.

The molecular gas in the vicinity of a galaxy nucleus is the raw material 
to support both star formation and (directly or indirectly) SMBH growth.
Conversely, either type of nuclear activity influences the chemical composition 
and excitation conditions of the surrounding molecular gas 
\citep[e.g.][]{sternberg94,kohno05,gracia-carpio06,krips08}. 
Studies of the molecular gas, therefore, can reveal important information 
about the relative importance of both phenomena. 
For example, the flux ratio between \coone\ and the dense gas
tracers HCN(1--0), and HCO$^+$(1-0) is a useful indicator for the
strength of an AGN \citep{kohno05}, with considerably higher
HCN/CO and HCO$^+$/CO flux ratios in AGN than in starbursts. This questions
the validity of the HCN intensity as a tracer for SF activity 
\citep{gao04a,gao04b,gracia-carpio06} in AGN sources. Models of
X-ray dominated regions have been evoked to explain the enhanced HCN
emission in AGN suggesting that the AGN itself is altering
the chemical properties of the surrounding ISM 
\citep[e.g.][]{meijerink05,meijerink06,meijerink07}

\begin{figure*}[ht]
\centering
\resizebox{0.95\hsize}{!}{\rotatebox{0}{\includegraphics{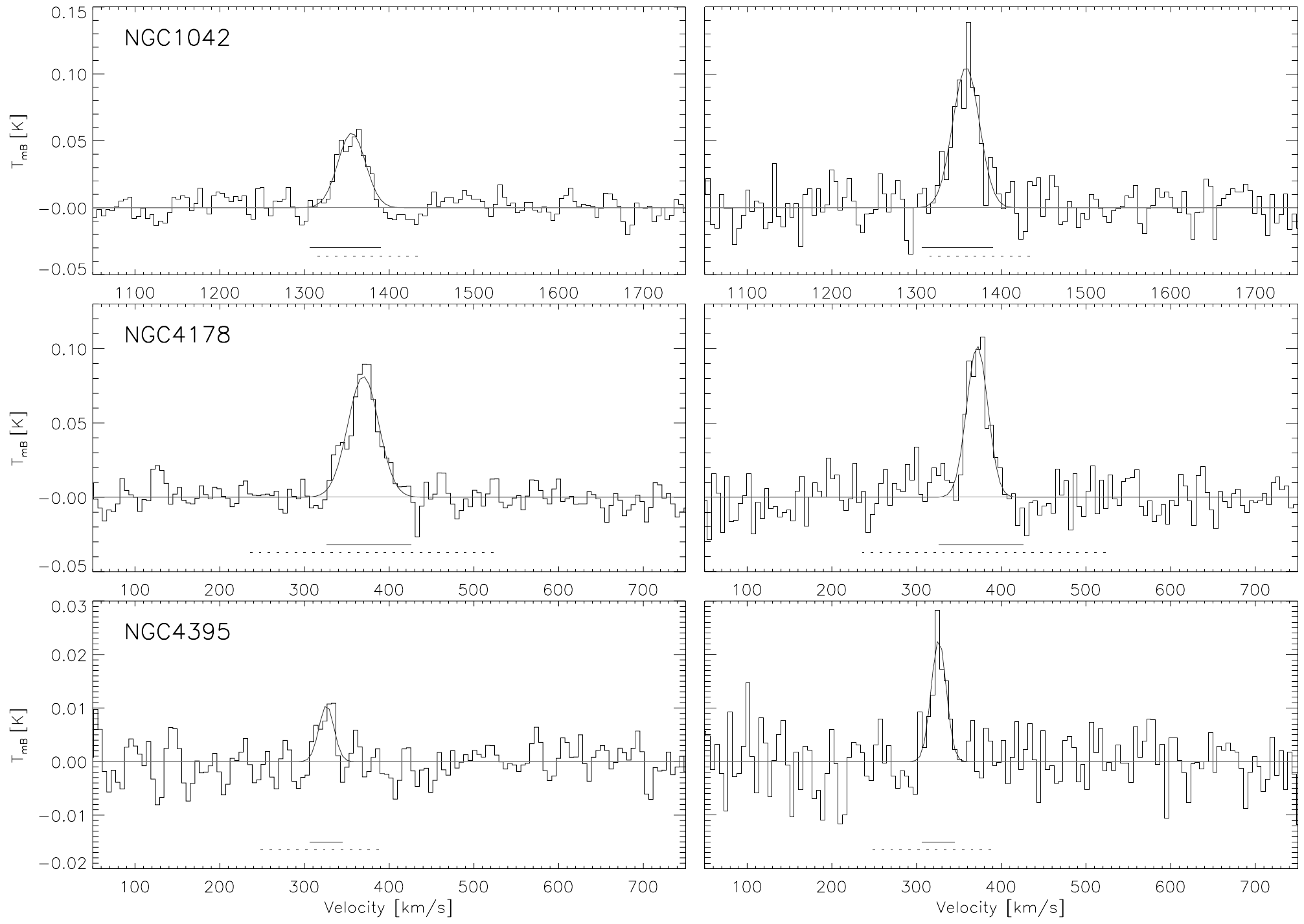}}} 
\caption{\coone\ (left) and \cotwo\ (right) emission line spectra for
three bulgeless spiral galaxies containing an AGN: NGC\,1042 (top), 
NGC\,4178 (middle), and NGC\,4395 (bottom). 
The velocity resolution for all spectra is $5.2\kms$. 
The horizontal lines in each panel indicate the HI line width at 20\% level 
($\rm W_{20}$, dashed) and the range over which the derived CO line intensities of 
Table~\ref{tab:lines} were integrated (solid). Also shown are Gaussian
fits to the line profiles which yield an alternative estimate for the line
intensities. 
}
\label{fig:spectra}
\end{figure*}

In this paper, we investigate the molecular gas content in all
three ``mini-AGN'' observable from the northern hemisphere, i.e. NGC\,1042,
NGC\,4178, and NGC\,4395, with the goal of comparing them
to more luminous AGN on the one hand, and inactive late-type
disks on the other. Some properties of the target galaxies and their AGN are 
summarized in Table\,\ref{tab:obs}. In addition to an SMBH, 
all three galaxies also host a prominent NSC, as can be seen in
Figure\,\ref{fig:mosaic}.

Following this introduction, we briefly describe the observations and data 
reduction procedure in \S\,\ref{sec:obs}, present and discuss our findings in
\S\,\ref{sec:results}, and summarize their implications
in \S\,\ref{sec:summary}.

\section{Observations and Data Reduction}
\label{sec:obs}

All data discussed here were obtained during the
period May 23-25, 2010 with the $30\m$
millimeter-wave telescope on Pico Veleta (Spain) operated by
the Institut de Radio Astronomie Millim\'etrique (IRAM). We used 
the EMIR receivers, with the autocorrelator WILMA as a backend.
This setup yields a resolution of $2\mhz$ and a bandwidth of $3.7\ghz$.
All observations were performed in wobbler-switching mode, with a throw 
in azimuth of 240\arcsec . The telescope pointing was checked on a nearby 
quasar about every 90 minutes. 

All observations were carried out in good weather conditions, the mean system 
temperatures being $320\kk$ on the T$_A ^*$ scale for the CO(1-0) and CO(2-1) 
transitions, and $105\kk$ for the low-frequency observations at $90\ghz$.
The main beam efficiencies were 0.81 ($90\ghz$), 0.77 ($115\ghz$), and 0.58 ($230\ghz$),
with half-power beam widths of about 28\arcsec\  ($90\ghz$), 21\arcsec\  ($115\ghz$),
and 11\arcsec\  ($230\ghz$). 

Data reduction was straightforward: the spectra
for each position were averaged. Only zero-order baselines
(i.e. continuum levels) were subtracted to obtain the final spectra.
We observed the central position of all three galaxies in the \coone\ and \cotwo\
transitions at $115\ghz$ and $230\ghz$ simultaneously, and in both polarizations.
In order to cover comparable physical areas in all three galaxies, we also carried 
out a 4-point mapping of NGC\,4395 in both CO lines, using offsets of $\pm$15\arcsec\ 
from the central position in both RA and Dec. 

In addition, we attempted to observe the lines of HCN(1-0) (rest frequency $88.631\ghz$), 
HCO+(1-0) ($89.189\ghz$), and HNC(1-0) ($90.979\ghz$) in NGC\,1042 and NGC\,4178.
For this purpose, we centered the receiver bandwidth at $89.1665\ghz$ and $89.5665\ghz$, 
respectively, to compensate for the recession velocity of the galaxies. 
Finally, we also targeted the isotopic line $^{13}$CO(2-1) of NGC\,4178 at $1\mm$.

\begin{table*}[h]
\centering
\caption{Target Galaxies \label{tab:obs}}
\begin{tabular}{lccccccc}
\hline
(1) & (2) & (3) & (4) & (5) & (6) & (7) & (8) \\
Galaxy & R.A. & Dec. & d & $\rm v_{sys}$ & $\rm W_{20}$ & $\rm L_{Bol}^{AGN}$ & $\rm M_{BH}$  \\
& & & Mpc & [km/s] & [km/s] & [erg/s] & [$\msun$] \\
\hline
  NGC\,1042  & 02h40m23.8s & -08:26:00.0  & 16.7 & 1377 & 124  & $8\cdot 10^{39}$  &  $\rm 60\lta M \lta 3\cdot10^6$ \\ 
  NGC\,4178  & 12h12m46.3s &  10:51:54.0  & 16.8 & 381 & 291  & $8\cdot 10^{41}$ & $\rm M \gta 6000$  \\
  NGC\,4395  & 12h25m48.8s &  33:32:48.0  & 4.1 & 318 & 140  & $5.4\cdot 10^{40}$ & $ (3.6\pm 1.1)\cdot 10^5$ \\
\end{tabular}
\tablefoot{Systemic velocity (Col. 4) and HI line width at 20\% level
(Col. 5) are from \cite{fisher81}. The distances to NGC\,1042 and 
NGC\,4178 are from \cite{tully88}, that to NGC\,4395 from \cite{thim04}. 
References for AGN properties and SMBH masses in Cols. (7) and (8) are as
follows: NGC\,1042: \cite{shields08}, NGC\,4178: \cite{satyapal09}, NGC\,4395: \cite{peterson05}.}
\end{table*}

\section{Results and Discussion}
\label{sec:results}
\subsection{Integrated Line Intensities und Upper Limits}
\label{subsec:limits}
The \coone\ and \cotwo\  spectra for the central position of all three
galaxies are presented in Figure~\ref{fig:spectra}. Both lines are clearly
detected in all three galaxies. We derive the velocity-integrated line 
intensities via two different methods: 
first, we simply sum up all channels with significant emission. We use the
same velocity window for \coone\ and \cotwo\ because the underlying
assumption is that both lines originate from the same gas distribution. The integration
windows are indicated by solid horizontal lines in Figure~\ref{fig:spectra}.

We also fit Gaussian profiles to the lines, which yields an alternative estimate
for the line intensities. These fits are also shown in Figure~\ref{fig:spectra}.
The agreement between the two methods is very good: the derived
line intensities agree to within 5\%.
In Table~\ref{tab:lines}, we list the average of the two methods.

A comparison with other CO observations in the literature yields the following:
NGC\,1042 was observed by \cite{braine93} who measure \ione $=2.9\pm0.3 \kkms$
with the IRAM $30\m$ telescope, in reasonable agreement with our results. 
NGC\,4178 was observed by \cite{kenney88}
who report a (marginal) detection of \ione $=0.9\pm0.3\kkms$. For the same galaxy,
\cite{boselli95} report \ione $=1.6\pm 0.5\kkms$, observed with the SEST $15\m$.
The fact that \ione\ is significantly smaller over the $43\as$ SEST beam than what
we measure with the IRAM telescope may indicate that the CO emission is less extended
than the SEST beam. On the other hand, it is surprising that \cite{komugi08} only 
give an upper limit of \ione $< 3.4\kkms$ over the $16\as$ beam of the Nobeyama $45\m$ 
telescope. Lastly, NGC\,4395 has, to our knowledge, not been detected in CO before.

\begin{table*}[h]
\centering
\caption{Summary of measured velocity-integrated line intensities and derived molecular gas masses. 
\label{tab:lines}
}
\begin{tabular}{lcccccc}
\hline
(1) & (2) & (3) & (4) & (5) & (6) & (7) \\
Galaxy & \ione & \itwo & \ihcn & \ihcoplus & \itco & \mhtwo \\
& [$\kkms$] & [$\kkms$] & [$\kkms$] & [$\kkms$] & [$\kkms$] & [$10^6\msun$] \\
\hline
NGC 1042          &  2.26$\pm$0.11  & 4.10$\pm$0.21 & $<$ 0.13 &    $<$ 0.11          & n.o.   & 27.0 \\
NGC 4178          &  3.78$\pm$0.13  & 3.19$\pm$0.18 & $<$ 0.10   & $<$ 0.07 &  $<$ 0.30 & 45.6 \\
NGC4395	(0,0)	  &  0.23$\pm$0.04  & 0.50$\pm$0.05   & n.o.    & n.o.   &  n.o. & 0.2 \\
NGC4395 (15,0)  &   $<$0.26  & $<$0.29   & n.o.    & n.o.   &  n.o. & $<$ 0.2 \\
NGC4395 (-15,0) &   $<$0.25  & $<$0.25   & n.o.    & n.o.   &  n.o. & $<$ 0.2 \\
NGC4395 (0,15)  &   $<$0.31  & $<$0.40   & n.o.    & n.o.   &  n.o. & $<$ 0.2 \\
NGC4395 (0,-15) &   $<$0.24  & $<$0.31  & n.o.    & n.o.   &  n.o. & $<$ 0.2 \\
\end{tabular}
\tablefoot{The entry n.o. stands for "not observed". Upper limits for the line intensities were derived as described 
in \S\,\ref{subsec:limits}.}
\end{table*}

In Figure\,\ref{fig:spectra}, we do not show the CO spectra for the off-nuclear pointings in NGC\,4395, because
none of them shows any significant CO emission. We list in Table\,\ref{tab:lines} conservative upper limits
for the \coone\ and \cotwo\ lines, derived as 
$ I_{\rm CO} < 3\cdot rms \cdot \sqrt{\Delta _{\rm v}\cdot \delta _{\rm v}} $ 
where $rms$ is the rms noise (per channel)
of the spectrum in question, $\Delta _{\rm v}$ is the integration window of the CO lines 
detected in the nucleus, and $\delta _{\rm v}$ is the velocity resolution, i.e. the channel width.

We also did not detect significant emission from HCN, HCO+(1-0), or $^{13}$CO(2-1)
in any of the (observed) nuclei, the corresponding 3$\sigma$ upper limits are reported in Cols. 4-6 of 
Table~\ref{tab:lines}. 
We also searched the entire receiver bandwidth of 3.7GHz for possible detections of 
other emission lines; no lines above the limits reported for the above lines were detected.

\subsection{Estimates of Molecular Gas Masses}
\label{subsec:masses}
We can estimate the mass of molecular hydrogen contained within the area covered by 
the IRAM beam by using the generic Galactic conversion factor of  
$X = 2.3 \cdot 10^{20} \cm^{-2}(\kkk)^{-1}$ \citep{strong88}, which yields:

\beq \label{eq:mh2}
M_{\rm H_2} = 97 \cdot D^2\cdot I_{10}\cdot \Theta^2 \>\>\>\> [\msun]
\eeq

when assuming a Gaussian beam with solid angle $1.13 \cdot \Theta^2$
\citep[for details, see][]{boker03}.
Here, $D$ is the galaxy distance measured in Mpc, $\Theta$ is the half power 
beam width of the 30\,m-telescope ($21\as$), and \ione \ the velocity integrated 
\coone\ intensity in $\kkms$.
The resulting \htwo\ masses for the three target galaxies are listed in Col.~7
of Table\,\ref{tab:lines}. The gas masses are rather low, especially in NGC\,4395
where the entire \htwo\ mass could be contained in a single giant molecular cloud 
(GMC) complex. 

This conclusion is independent of the use of the generic conversion 
factor, because the galaxies studied here are neither extremely faint, nor 
have extreme metallicities. For example, the 
oxygen abundance in the disk of NGC\,4395, the faintest of our targets, has been 
reported by \cite{pilyugin01} to be $12+log(O/H)=8.3-8.4$, or about one third
of the solar value \citep[8.76, ][]{caffau08}. 
Assuming that the nuclear region has at least this abundance, using the 
metallicity-dependent conversion factor proposed by \cite{boselli02}
would increase the derived gas mass by no more than a factor of 2.5.

\subsection{CO line ratios}
\label{subsec:ratios}
A number of studies \citep[e.g.][]{braine93,aalto95,papadopoulos98} 
have investigated the $r_{21}\equiv I_{21}/I_{10}$ ratio in
galaxies, with the aim to extract information about the optical depth and 
excitation state of the gas. All these studies find that in the vast majority
of galaxies, the value of $r_{21}$ is unity, with a relatively small scatter.

For example, \cite{braine93} find an average
value of $r_{21}=0.89\pm0.06$ for a sample of 81 nearby spiral
galaxies. Similarly, \cite{aalto95} find a mean value of $r_{21}=0.93$
in a sample of 32 starburst galaxies. Taken at face value, these
results suggest that most of the molecular gas in the central regions
of galaxies is optically thick and thermally excited. 

However, because line intensities measured on the main beam temperature
scale $\rm T_{mb}$ are equivalent to a surface brightness, and the two lines
are observed with different telescope beams, interpreting the value of
$r_{21}$ is difficult if the source size is unknown \citep{papadopoulos98}.

\begin{figure*}[ht]
\centering
\resizebox{0.46\hsize}{!}{\rotatebox{0}{\includegraphics{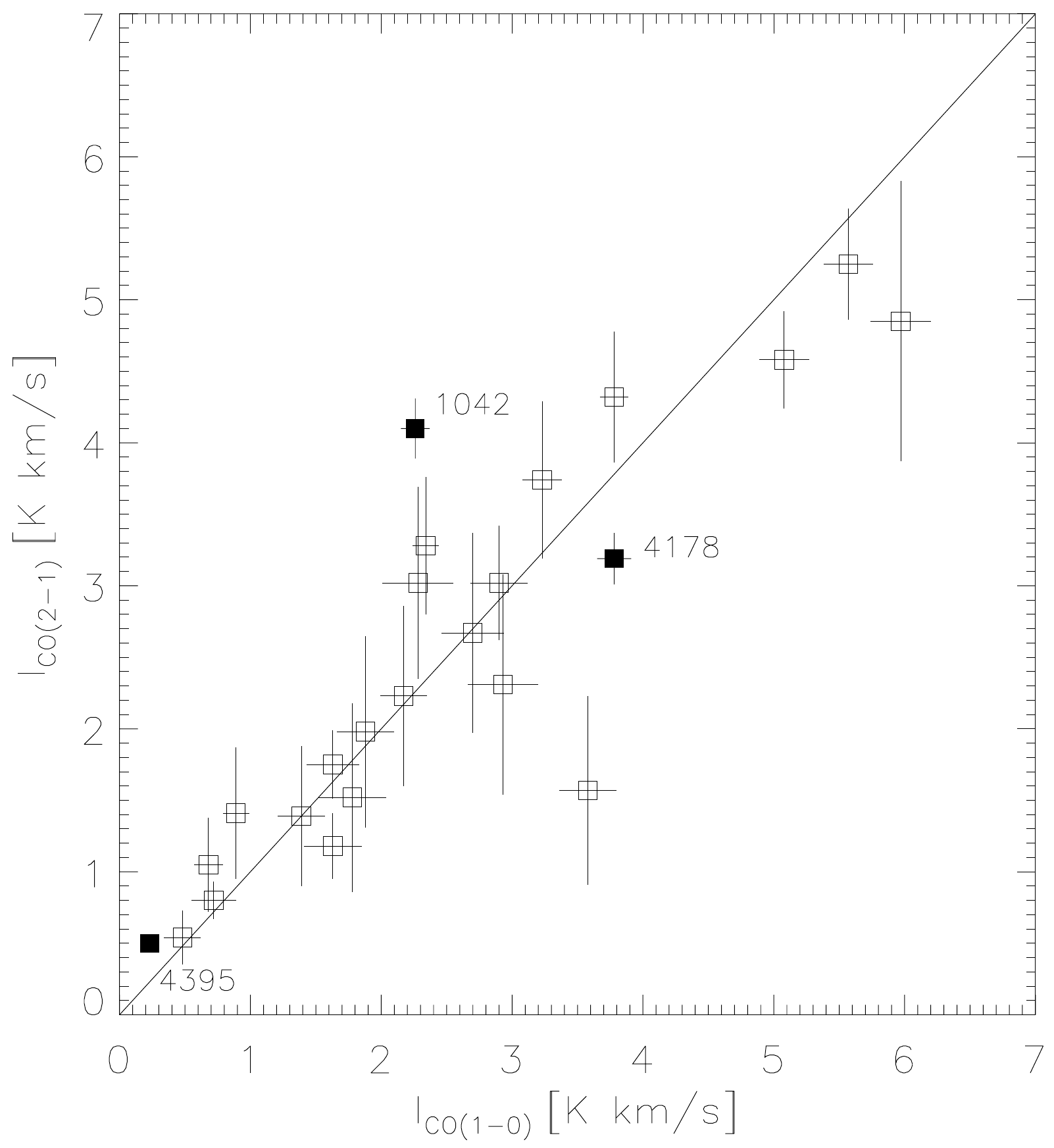}}}
\resizebox{0.48\hsize}{!}{\rotatebox{0}{\includegraphics{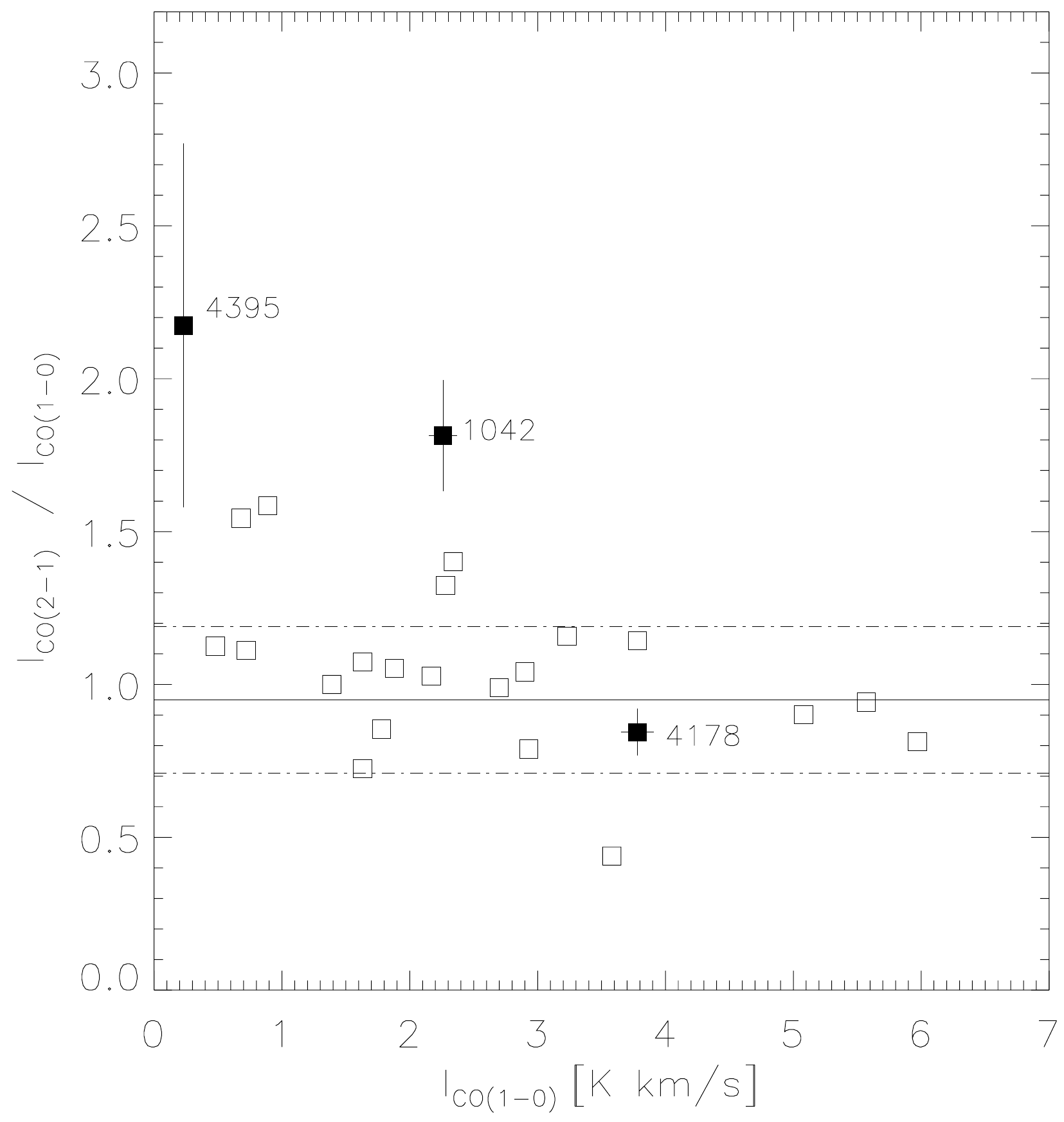}}} 
\caption{Left: line intensities of \coone\ and \cotwo\ for quiescent late-type
spirals \citep[][open squares]{boker03} and those with low-luminosity 
AGN (filled squares, this work). 
Right: $I_{21}/I_{10}$ ratio vs. $I_{10}$, indicating the elevated line ratio for
NGC\,1042 and NGC\,4395. Here, the error bars for the comparison sample are 
not shown for clarity. The horizontal lines indicate the mean 
value of $I_{21}/I_{10}$ (solid) and the mean $\pm 1\sigma$ (dashed) for the
comparison sample.}
\label{fig:ratios}
\end{figure*}

In Fig.\,\ref{fig:ratios}, we compare the line intensities of the \coone\ and
\cotwo\ lines in the target AGN to those in a comparison sample of morphologically
similar, but quiescent, late-type spirals at similar distances and with moderate
inclinations. This comparison sample was observed with the same telescope, 
and analyzed in the same way as the data presented here \citep{boker03}. 

As can be seen in the right panel, the value of $r_{21}$ appears unusually
high in two of the AGN (NGC\,1042 and NGC\,4395), while it is rather 
typical in NGC\,4178. In this context, it is important to note that both NGC\,1042 and
NGC\,4395 are oriented nearly face on (as is the entire comparison sample), 
while NGC\,4178 is an edge-on spiral (see Fig.\,\ref{fig:mosaic}). 
Therefore, it is possible (and in fact rather likely) 
that the line-of-sight towards the nucleus of NGC\,4178 contains gas located in 
its outer disk, and that our measurements for this galaxy are not really representative 
of the {\it nuclear} gas properties. 

In order to quantify the significance level of the elevated $r_{21}$ values in
the three AGN, we performed a two-Sample t-Test. This test is designed
to check whether the null hypothesis (i.e. that the filled squares in Fig.\,\ref{fig:ratios}
are randomly drawn from the same population as the open squares) can be rejected 
with confidence. The mean $r_{21}$ in NGC\,1042 and NGC\,4395 differs from
the mean $r_{21}$ of the comparison sample by 4.75 standard errors, and the probability
that both are part of the same parent population is 0.0001. In other words, it
can be concluded with 99.99\% confidence that the two face-on AGN discussed in
this paper indeed have higher $r_{21}$ values than their non-active 
counterparts\footnote{For completeness, we mention that the significance
is reduced to 99\% if the edge-on AGN NGC\,4178 is included in the analysis.}.

Of course, given that there are only two objects studied so far, one cannot
conclude that all low-luminosity AGN observed face-on show elevated CO line ratios. 
Nevertheless, Fig.\,\ref{fig:ratios} provides tentative evidence that the $r_{21}$ 
value may be a good indicator for low-level nuclear activity. The reliability of 
$r_{21}$  in this context should be tested with similar observations of other mini-AGNs.
An obvious candidate for such an observational test is NGC\,3621, the only other 
bulge-less spiral galaxy with an AGN, which, unfortunately, cannot be observed 
from the northern hemisphere, but should become a prime target for ALMA.

\subsection{Effects of Source Size}
\label{sec:exc}
We now discuss some implications of the fact that we do not know a priori the spatial extent
of the CO-emitting region. In principle, one can estimate the intrinsic source size using the 
following relation between the absolute 
source brightness $T_{\rm s}$ and the main beam brightness 
$T_{\rm b}$: $T_{\rm s} = T_{\rm b} \frac{\theta_{\rm s}^2 + \theta_{\rm b}^2}{\theta_{\rm s}^2}$,
where $\theta_{\rm s}$ is the source size and $\theta_{\rm b}$ the beam size.

In order to solve for the source size, we can assume for the moment that the intrinsic 
temperatures of the CO(1-0) and CO(2-1) line emission are  the same, i.e. that $\rm T_{\rm s10} = T_{\rm s21}$,
which leads to:
\begin{equation}
\theta_{\rm s} = \sqrt{(T_{\rm b21}\cdot \theta_{\rm b21}^2 - T_{\rm b10}\cdot \theta_{\rm b10}^2)/(T_{\rm b10} - T_{\rm b21})}
\end{equation}
Assuming also that the line profiles of the \coone\ and \cotwo\ are identical, one can replace the
peak temperatures with the less noisy integrated line intensities from Table\,\ref{tab:lines}. 
This results in source sizes of $16\as$ and $12\as$ for NGC\,1042 and NGC\,4395, respectively. 
Thus, in order to explain the observed elevated $r_{21}$ value solely by beam dilution effects, the 
CO emission cannot be more extended than this. 

Unfortunately, the sensitivity of our off-nuclear spectra does not allow to place strong constraints on the actual
extent of the CO emission, because the upper limits for the line intensities (see Table 2) are still consistent
with extended emission. We therefore use the Spitzer/IRAC $8\mum$ image of NGC\,4395 \citep{dale09} to check 
whether this a compact CO source is a reasonable assumption. Because the IRAC $8\mum$ channel is dominated 
by emission from 
large molecules, mainly Polycyclic Aromatic Hydrocarbonates, it should be a good tracer of molecular gas.
Inspection of this image shows a rather compact emission region of $\approx 11\as$ ($220\pc$)
diameter surrounding the central point source. However, there is also a network of narrow and 
well-defined dust filaments throughout the central $1\am$ which could also contribute to the CO flux.

On the other hand, {\it none} of the galaxies in the comparison sample {\it without} AGN shows $r_{21}$ 
values as high as those observed in NGC\,1042 and NGC\,4395, even though they are morphologically 
similar and at comparable distances. Interestingly, there are four galaxies (NGC\,2805, NGC\,3445, NGC\,3782,
and NGC\,5669) that fall above the $1\sigma$ deviation. These may be candidates for 
follow-up searches for signs of nuclear activity, for example with deep optical spectroscopy that is 
able to accurately measure emission line ratios in the nucleus, similar to the case of NGC\,1042 \citep{shields08}.

Because there is no apparent reason why the gas distribution in NGC\,4395 and NGC\,1042 should be 
more concentrated than in the galaxies of the comparison sample, we consider it unlikely (but cannot rule out) 
that the elevated $r_{21}$ ratio is explained by a very compact CO source in NGC\,4395 and NGC\,1042. 
Instead, we suggest that the molecular gas in these two objects is {\it not} in thermal equilibrium, 
and that the CO(2-1) emission is excited by ionizing radiation. 

This ionizing radiation can, in principle, be produced either by an AGN or by young massive stars in 
the nuclear star cluster. However, given that most galaxies in the comparison sample also have a 
nuclear cluster, but do not show elevated values of $r_{21}$, it seems plausible to assume that it 
is indeed the AGN that is responsible for the excitation of CO(2-1) in these two galaxies. 

\subsection{Dense Molecular Gas around weak AGN?}
\label{subsec:dense}

\begin{figure}[ht]
\centering
\resizebox{0.9\hsize}{!}{\rotatebox{0}{\includegraphics{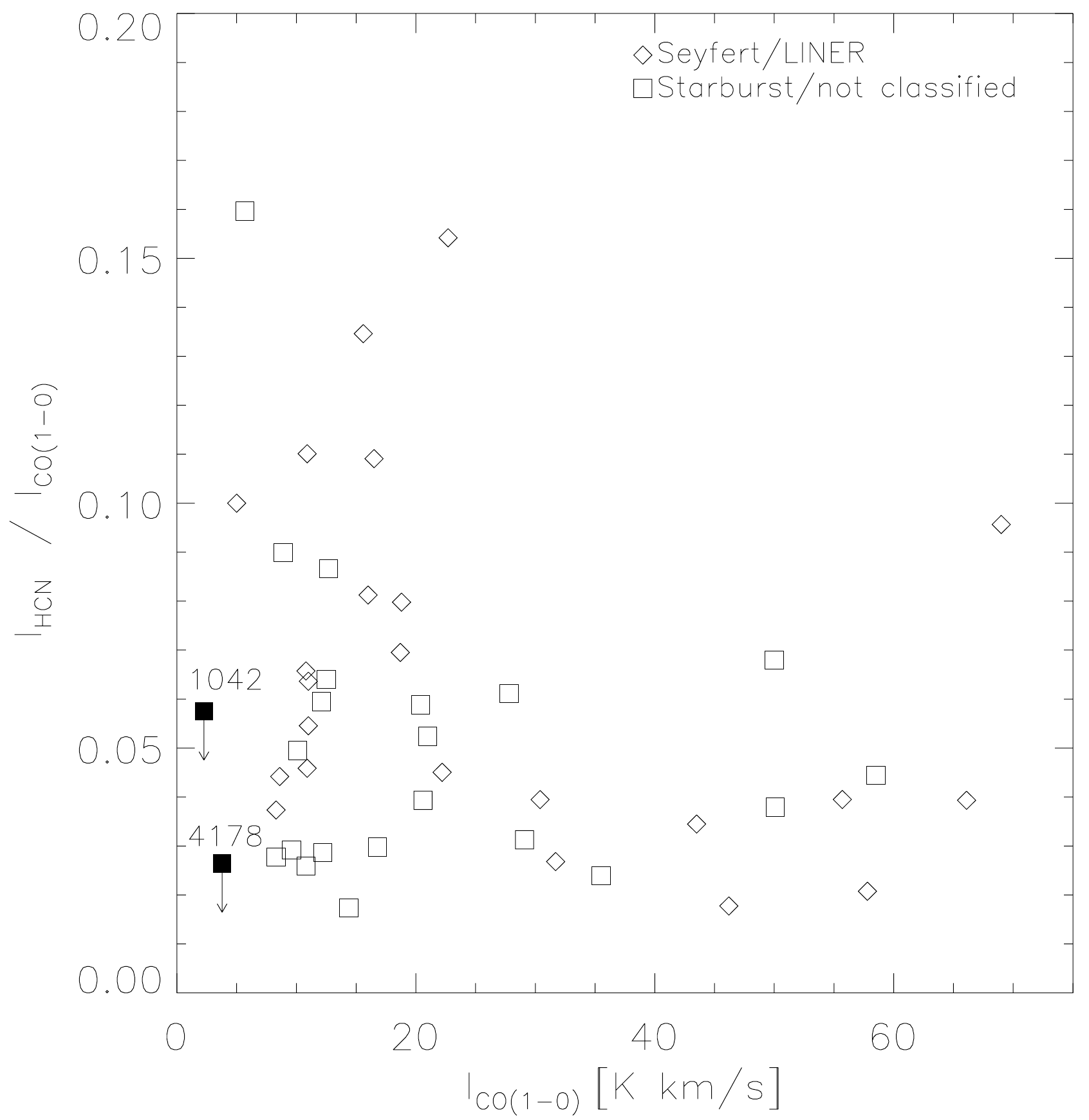}}} 
\caption{Comparison of the $I_{\rm HCN} / I_{10}$ ratio in NGC\,1042 and
NGC\,4395 to that in a sample of active (diamonds) and non-active (squares)
galaxies from \cite{gao04b}. 
}
\label{fig:hcnvsco}
\end{figure}

Because the critical density for the excitation of the ground-level of the HCN
molecule is $\rm n_{cr} \approx 10^5 cm^{-3}$, the HCN(1-0) line  
probes significantly higher gas densities than the $^{12}$CO(1-0) line. 
In principle, the value of $r_{\rm HCN}\equiv $\ihcn /\ione\ can thus be used to estimate the 
average density of the molecular gas observed. This has recently been 
investigated in more detail by \cite{matsushita10} who studied the 
behavior of $r_{\rm HCN}$ as a function of kinetic temperature and
$\rm H_2$ density by utilizing a Large Velocity Gradient (LVG) model.
Their Fig. 5 shows that $r_{\rm HCN}$ is very sensitive to
the density of the gas while being insensitive to the actual kinetic
temperature. Our observed upper limits for $r_{\rm HCN}$ of 0.058 and 0.026 
for NGC\,1042 and NGC\,4178 indicate gas densities below $\rm
1-2\times10^3\,cm^{-3}$. 

Our derived upper limits for $r_{\rm HCN}$ in NGC\,1042 and NGC\,4178 lie 
at the low end of the observed range of $r_{\rm HCN}$, but are consistent 
with both starforming galaxies and low-luminosity AGN. 
This is illustrated in Figure\,\ref{fig:hcnvsco} which compares the upper 
limits for these two galaxies to the measurements of \cite{gao04b} 
who studied a sample of 53 galaxies of various
types, including both active and non-active nuclei.

Taken together, this suggests that the molecular ISM in the two low-luminosity AGN
studied here is similar to that in other galaxies, and certainly shows no evidence for
an enhanced gas density as observed in some nearby AGN \citep{kohno08,juneau09}. 
Hence, we conclude that the presence of a weak AGN in NGC\,1042 and NGC\,4178 
has no impact on the observed ISM density. 

\section{Summary}
\label{sec:summary}

We have measured the \coone\ and \cotwo\ emission from the nuclei of three bulgeless
spiral galaxies with low-luminosity AGN: NGC\,1042, NGC\,4178, and NGC\,4395. In the case
of NGC\,4395, this constitutes the first reported detection of CO. The CO emission is rather 
faint, and the inferred molecular gas masses are low. This is especially true for NGC\,4395 
where the entire \htwo\ mass could be contained in a single GMC.

Two of the three target galaxies (NGC\,1042 and NGC\,4395) show elevated values
of the intensity ratio $r_{21}$. Both of these are oriented nearly face-on, and thus offer an 
unobstructed sight line into their nucleus. The third target (NGC\,4178),
in contrast, is highly inclined, and the fact that it shows a value of $r_{21}$ that is typical for other 
late-type spirals without an AGN may therefore be explained by its CO spectra being dominated 
by gas located in the outer disk. 

We have discussed, even though we do not have a priori knowledge of the size of the CO-emitting
region in our targets, that the elevated $r_{21}$ values in the two face-on AGNs are unlikely 
caused by beam dilution due to an extremely compact source size. This is based on estimates of 
the required source size that could explain the observed line ratios and comparison with the 
gas morphology deduced from Spitzer imaging. Instead, we argue that the elevated $r_{21}$ 
values are more likely caused by ionizing radiation, implying that even energetically weak AGN 
can impact the physical state of the surrounding gas. 

We do not detect any tracers of dense molecular gas such as HCN or HCO+, but the sensitivity
of our observations allows us to establish meaningful upper limits. The derived gas density is less than 
$n_{\rm H_2}\approx 2\cdot 10^3\,\rm cm^{-3}$ which is at the low end of the range observed in 
nearby quiescent galaxies, and certainly lower than in more energetic AGN. 
This implies that using the value of $R_{\rm HCN}$ as an indicator for the presence
of an (obscured) AGN is not reliable for weak AGN. More generally, the scarcity of dense 
gas in our target galaxies suggests that the conditions for star formation 
are poor in these nuclei, and may well be linked to their inefficient bulge growth.

Our observations have demonstrated that studies of the molecular gas around the faintest AGN
are feasible, but at the limit of even the best current observatories. Looking ahead, 
however, these and similar galaxies should be interesting targets for future interferometers,
and in particular the ALMA observatory, because its improved sensitivity and spatial resolution 
promises to reveal in detail the feedback mechanism(s) between the AGN, the nuclear star cluster,
and the ISM in their vicinity.

\acknowledgements
We are grateful for a thoughtful referee report by Jeremy Lim, whose comments prompted a more 
thorough statistical analysis of our results, and helped to improve the overall quality of the paper.
We are also grateful to S. Meidt and P. Jakobsen for help with identifying the correct statistical test
to gauge the significance of our results. 
This work is based on observations with the Instituto de Radioastronomia Milim\'etrica (IRAM) $30\m$ telescope.
IRAM is supported by INSU/CNRS (France), MPG (Germany, and IGN (Spain).
We made use of the Nasa Extragalactic Database (NED) and of the Lyon Extragalactic Database (LEDA).
UL acknowledges support by the research project   AYA2007-67625-C02-02  from the Spanish Ministerio de
Ciencia y Educaci\'on and the Junta de Andaluc\'\i a (Spain) grant FQM-0108.
UL warmly thanks IPAC (Caltech) for their hospitality, where this work was finished during a sabbatical stay.

\bibliographystyle{aa} 
\bibliography{IRAM_miniAGN} 
\end{document}